\documentclass[prd,aps,showpacs,nofootinbib,preprint]{revtex4}
\usepackage{graphicx,color,amsmath,amsxtra}
\usepackage{epsf}
\usepackage{amssymb}
\usepackage{enumerate}
\usepackage{hhline}
\usepackage{array}
\usepackage{tabularx}
\usepackage{hangcaption}

\newcommand{\bear}{\begin{array}}  \newcommand{\eear}{\end{array}}
\newcommand{\bea}{\begin{eqnarray}}  \newcommand{\eea}{\end{eqnarray}}
\newcommand{\beq}{\begin{equation}}  \newcommand{\eeq}{\end{equation}}
\newcommand{\bef}{\begin{figure}}  \newcommand{\eef}{\end{figure}}
\newcommand{\bec}{\begin{center}}  \newcommand{\eec}{\end{center}}

\newcommand{\Eqn}[1]{&\hspace{-0.2em}#1\hspace{-0.2em}&}
\def\Vec#1{\mbox{\boldmath $#1$}}

\def\be{\begin{equation}}
\def\ee{\end{equation}}
\def\bea{\begin{eqnarray}}
\def\eea{\end{eqnarray}}
\def\beq{\begin{eqnarray}}
\def\eeq{\end{eqnarray}}

\def\be{\begin{equation}}
\def\ee{\end{equation}}
\def\bea{\begin{eqnarray}}
\def\eea{\end{eqnarray}}
\def\beq{\begin{eqnarray}}
\def\eeq{\end{eqnarray}}

\begin{document}

\title{Generic feature of future crossing of phantom divide 
in viable $f(R)$ gravity models
}

\author{Kazuharu Bamba\footnote{E-mail address: 
bamba@phys.nthu.edu.tw}, 
Chao-Qiang Geng\footnote{E-mail address: geng@phys.nthu.edu.tw} 
and 
Chung-Chi Lee\footnote{E-mail address: g9522545@oz.nthu.edu.tw} 
}
\affiliation{
Department of Physics, National Tsing Hua University, Hsinchu, Taiwan 300 
}

%\date{\today}

%%%%%%%%%%%%%%%%%%%%%
%  Abstract
%%%%%%%%%%%%%%%%%%%%%
\begin{abstract}

We study the equation of state for dark energy 
and explicitly demonstrate that the future crossings of the phantom 
divide line $w_{\mathrm{DE}}=-1$ are the generic feature in the existing 
viable $f(R)$ gravity models. 
We also explore the future evolution of the cosmological horizon entropy 
and illustrate that the cosmological horizon entropy oscillates with time 
due to the oscillatory behavior of the Hubble parameter. 
The important cosmological consequence is that in the future, 
the sign of the time derivative of the Hubble parameter changes 
from negative to positive in these viable $f(R)$ gravity models. 

\end{abstract}
%%%%%%%%%%%%%%%%%%%%%

%----------------------------
%\pacs{
%Keywords:
%}
%\preprint{}
%\hspace{13.0cm}
%----------------------------

\maketitle
%==============================================================================

%%%%%%%%%%%%%%%%%%%%%%%%%%%
%%%  Sec. I
%%%%%%%%%%%%%%%%%%%%%%%%%%%
\section{Introduction}

The cosmological observations such as supernovae Ia~\cite{SN1},
cosmic microwave background radiation~\cite{WMAP}, 
large scale structure~\cite{LSS}, and weak lensing~\cite{WL}
have revealed that the 
universe has been undergoing an  accelerating expansion 
since the recent ``past'', which is one of  the most challenging problems in physics today. 
There are two representative approaches to explain the late time 
acceleration of the universe. 
One is the introduction of ``dark energy'' in the framework of general 
relativity~\cite{Copeland:2006wr}. The other is the consideration of a modified gravitational theory, 
such as $f(R)$ gravity~\cite{Nojiri:2006ri, Sotiriou:2008rp, DeFelice:2010aj}. 
In this study, we will concentrate on the later approach. 

It has been commonly adopted that a viable $f(R)$ gravity model needs to 
satisfy the following conditions: 
(a) positivity of the effective gravitational coupling, 
(b) stability of cosmological perturbations~\cite{Nojiri:2003ft}, 
(c) asymptotic behavior to the standard $\Lambda$-Cold-Dark-Matter 
($\Lambda\mathrm{CDM}$) model in the large curvature regime, 
(d) stability of the late-time de Sitter point~\cite{Muller:1987hp}, 
(e) constraints from the equivalence principle, 
and 
(f) solar-system constraints~\cite{Solar-System-Constraints}. 
%
%%%
Several viable models 
have been constructed in the literature, 
such as the popular ones: 
(i) Hu-Sawicki~\cite{Hu:2007nk}, 
(ii) Starobinsky~\cite{Starobinsky:2007hu}, 
(iii) Tsujikawa~\cite{Tsujikawa:2007xu}, 
and (iv) the exponential gravity~\cite{Exponential-type-f(R)-gravity, 
Cognola:2007zu, Linder:2009jz,Bamba:2010ws} models with the explicit forms shown in Table~\ref{Table}. 
\begin{table}[htbp]
\caption{Explicit forms of $f(R)$ in 
(i) Hu-Sawicki, 
(ii) Starobinsky, 
(iii) Tsujikawa, 
and (iv) the exponential gravity models.}
\vskip 0.2in
\label{Table}
\begin{tabular}{|c||c|c|} \hline
 model & $f(R)$ & Constant parameters
\\ \hline \hline
(i) & $ R - 
\frac{c_1 R_{\mathrm{HS}} \left(R/R_{\mathrm{HS}}\right)^p}{c_2 
\left(R/R_{\mathrm{HS}}\right)^p + 1}$ &
$c_1$, $c_2$, $p(>0)$, $R_{\mathrm{HS}}(>0)$
\\ \hline
(ii) & $ R + 
\lambda R_{\mathrm{S}} \left[
\left(1+\frac{R^2}{R_{\mathrm{S}}^2} \right)^{-n}-1 
\right]$ & $\lambda (>0)$, $n (>0)$, $R_{\mathrm{S}}$
\\ \hline
%(iii) &  $\frac{R}{2} + \frac{
%R_{\mathrm{vac}}}{2\left(b+\ln\left(2\cosh\,b\right)\right)} 
%\ln \left[ \frac{\cosh\,\left(R/\epsilon_{\mathrm{AB}}-b\right)}{\cosh\,b}\right]$
%& $b$, $R_{\mathrm{vac}}$
%\\ \hline
(iii) &
$R - \mu R_{\mathrm{T}} 
\tanh\left( \frac{R}{R_{\mathrm{T}}} \right)$
&
$\mu (>0)$, $R_{\mathrm{T}} (>0)$
\\ \hline
(iv) &
$R -\beta R_{\mathrm{E}}\left(1-e^{-R/R_{\mathrm{E}}} 
\right)$
& $\beta$, $R_{\mathrm{E}}$
\\ \hline
\end{tabular}
%\label{Table}
\end{table}
For other viable models 
%%%
(e.g., models in Ref.~\cite{Nojiri-Odintsov}) 
%%%
and references, see a recent 
review in Ref.~\cite{DeFelice:2010aj}. 

Recently, 
the cosmological observational data~\cite{observational status} also seems to 
indicate the crossing of the phantom divide $w_{\mathrm{DE}}=-1$ of the 
equation of state for dark energy 
in the near ``past''. To understand such a crossing, many attempts have been 
made. 
The most noticeable one is to use a phantom field with a negative kinetic 
energy term~\cite{Phantom}. 
Clearly, it surfers a serious problem as it is not stable at the quantum 
level. 
On the other hand, 
the crossing of the phantom divide can also be realized in 
the above viable $f(R)$ 
models~\cite{Hu:2007nk, 
Linder:2009jz, Bamba:2010ws, 
Martinelli:2009ek} 
without violating any stability conditions.
This is probably 
the most peculiar character of the modified gravitational models. 
Other $f(R)$ gravity models  with realizing 
a crossing~\cite{Nojiri:2006ri, Abdalla:2004sw} 
as well as multiple crossings~\cite{Bamba:2009kc} of 
the phantom boundary have also been examined. 

However, most of the studies in $f(R)$ gravity 
have been focused on the past. 
In this paper, 
we would like to explore the future behaviors of the universe. 
In particular, 
we show that the viable $f(R)$ models 
generally exhibit 
the crossings of the phantom divide in the ``future'' too. 
In addition, we investigate the future evolution of the cosmological horizon 
entropy and demonstrate that the cosmological horizon entropy oscillates with 
time due to the oscillation of the Hubble parameter. 
%%%%% Units %%%%%

We use units of 
$k_\mathrm{B} = c = \hbar = 1$ and 
denote the gravitational constant $8 \pi G$ by 
${\kappa}^2 \equiv 8\pi/{M_{\mathrm{Pl}}}^2$ 
with the Planck mass of 
$M_{\mathrm{Pl}} = G^{-1/2} = 1.2 \times 10^{19}$\,\,GeV.
We assume the flat Friedmann-Lema\^{i}tre-Robertson-Walker (FLRW) 
space-time with the metric, 
\begin{eqnarray}
{ds}^2 = -{dt}^2 + a^2(t)d{\Vec{x}}^2\,,
\label{eq:3.1}
\end{eqnarray}
where $a(t)$ is the scale factor. 
%%%%%%%%%%%%%%%%%

%%%%%
The paper is organized as follows. 
In Sec.\ II, we explain $f(R)$ gravity and derive the gravitational field 
equations. 
We investigate the cosmological evolution in Sec.\ III. 
Finally, conclusions are given in Sec.\ IV. 
%%%%%

%%%%%%%%%%%%%%%%%%%%%%%%%%%
%%%  Sec. II
%%%%%%%%%%%%%%%%%%%%%%%%%%%
\section{$f(R)$ gravity}

The action of $f(R)$ gravity with matter is given by
\begin{equation}
I = \int d^4 x \sqrt{-g} \frac{f(R)}{2\kappa^2} + I_{\mathrm{matter}} 
(g_{\mu\nu}, \Upsilon_{\mathrm{matter}})\,,
\label{eq:1}
\end{equation} 
where $g$ is the determinant of the metric tensor $g_{\mu\nu}$, 
$I_{\mathrm{matter}}$ is the action of matter which is assumed to be minimally 
coupled to gravity, i.e., the action $I$ is written in the Jordan frame, 
and $\Upsilon_{\mathrm{matter}}$ denotes matter fields. 
Here, we use the standard metric formalism. 
%%%%%
By taking the variation of the action in Eq.~(\ref{eq:1}) with respect to 
$g_{\mu\nu}$, one obtains~\cite{Sotiriou:2008rp} 
\begin{equation}
F G_{\mu\nu} 
= 
\kappa^2 T^{(\mathrm{matter})}_{\mu \nu} 
-\frac{1}{2} g_{\mu \nu} \left( FR - f \right)
+ \nabla_{\mu}\nabla_{\nu}F -g_{\mu \nu} \Box F\,,
\label{eq:2}
\end{equation}
where 
$G_{\mu\nu}=R_{\mu\nu}-\left(1/2\right)g_{\mu\nu}R$ is 
the Einstein tensor, 
$F(R) \equiv d f(R)/dR$, 
${\nabla}_{\mu}$ is the covariant derivative operator associated with 
$g_{\mu \nu}$, 
$\Box \equiv g^{\mu \nu} {\nabla}_{\mu} {\nabla}_{\nu}$
is the covariant d'Alembertian for a scalar field, 
and 
$T^{(\mathrm{matter})}_{\mu \nu}$ 
is the contribution to the energy-momentum tensor from all 
perfect fluids of matter. 
%
%%%%%%%%%%%%%%%%%%%
%%%  Sec. III
%%%%%%%%%%%%%%%%%%%
%\section{Cosmological evolution}
%
{}From Eq.~(\ref{eq:2}), 
we obtain the following gravitational field equations: 
\begin{eqnarray} 
3FH^2 
\Eqn{=}
\kappa^2 \rho_{\mathrm{M}} +\frac{1}{2} \left( FR - f \right) 
-3H\dot{F}\,,
\label{eq:3.2} \\ 
-2F\dot{H}  
\Eqn{=}
\kappa^2 \left( \rho_{\mathrm{M}} + P_{\mathrm{M}} \right)
+\ddot{F}-H\dot{F}\,,
\label{eq:3.3}
\end{eqnarray} 
where $H=\dot{a}/a$ is the Hubble parameter, 
the dot denotes the time derivative of $\partial/\partial t$, and 
$\rho_{\mathrm{M}}$ and $P_{\mathrm{M}}$ are 
the energy density and pressure of all perfect fluids of matter, 
respectively.

%%%%%%%%%%%%%%%%%%%
%%%  Sec. III
%%%%%%%%%%%%%%%%%%%
\section{Cosmological evolution}

We analyze the cosmological evolution of dark energy 
including those of non-relativistic matter (cold dark matter and baryon) and 
radiation by solving Eq.~(\ref{eq:3.2}) 
and $R=6\left(2H^2+\dot{H}\right)$. 
In particular, we examine the equation of state for dark energy, given by
\begin{eqnarray} 
w_{\mathrm{DE}} \Eqn{\equiv} P_{\mathrm{DE}}/\rho_{\mathrm{DE}},
\nonumber\\
 \rho_{\mathrm{DE}} \Eqn{=} \frac{1}{\kappa^2} \left[ 
\frac{1}{2} \left( FR - f \right) 
-3H \dot{F} 
+3\left(1-F\right)H^2 
\right]\,,
 \nonumber\\
 P_{\mathrm{DE}} \Eqn{=} \frac{1}{\kappa^2} 
\left[ 
-\frac{1}{2} \left( FR - f \right) 
+\ddot{F}+2H \dot{F}
-\left(1-F\right)\left(2\dot{H}+3H^2\right) 
\right]\,,
 \label{eq:W}
\end{eqnarray}
in the future, 
i.e., 
the region with the redshift $z<0$. 
Furthermore, we investigate the evolution of the cosmological horizon entropy. 
It has been shown that it is possible to obtain a picture of equilibrium 
thermodynamics on the apparent horizon in the FLRW background 
for $f(R)$ gravity 
due to a suitable redefinition of an energy momentum tensor of 
the ``dark'' component that respects a local energy conservation 
in Ref.~\cite{Bamba:2009id}. 
%%%%%
In this picture, the horizon entropy 
$S=A/(4G)$~\cite{Bardeen:1973gs} 
where $A$ is the area of the apparent horizon, 
is simply expressed 
as~\cite{Bamba:2009id} 
\begin{equation}
S=\frac{\pi}{GH^2}\,.
\label{eq:A-A-1}
\end{equation}
We note that 
in the context of modified gravity theories including $f(R)$ gravity,
a horizon entropy $\hat{S}$ associated with a Noether 
charge has been proposed by Wald~\cite{Wald entropy}, 
given by~\cite{Jacobson:1993vj} 
$\hat{S}=F(R)A/\left(4G\right)$. 
%%%%%

%%%%%% Fig. 1 %%%%%%%%%
\begin{center}
\begin{figure}[tbp]
\begin{tabular}{ll}
\begin{minipage}{80mm}
\begin{center}
\unitlength=1mm
\resizebox{!}{6.5cm}{
   \includegraphics{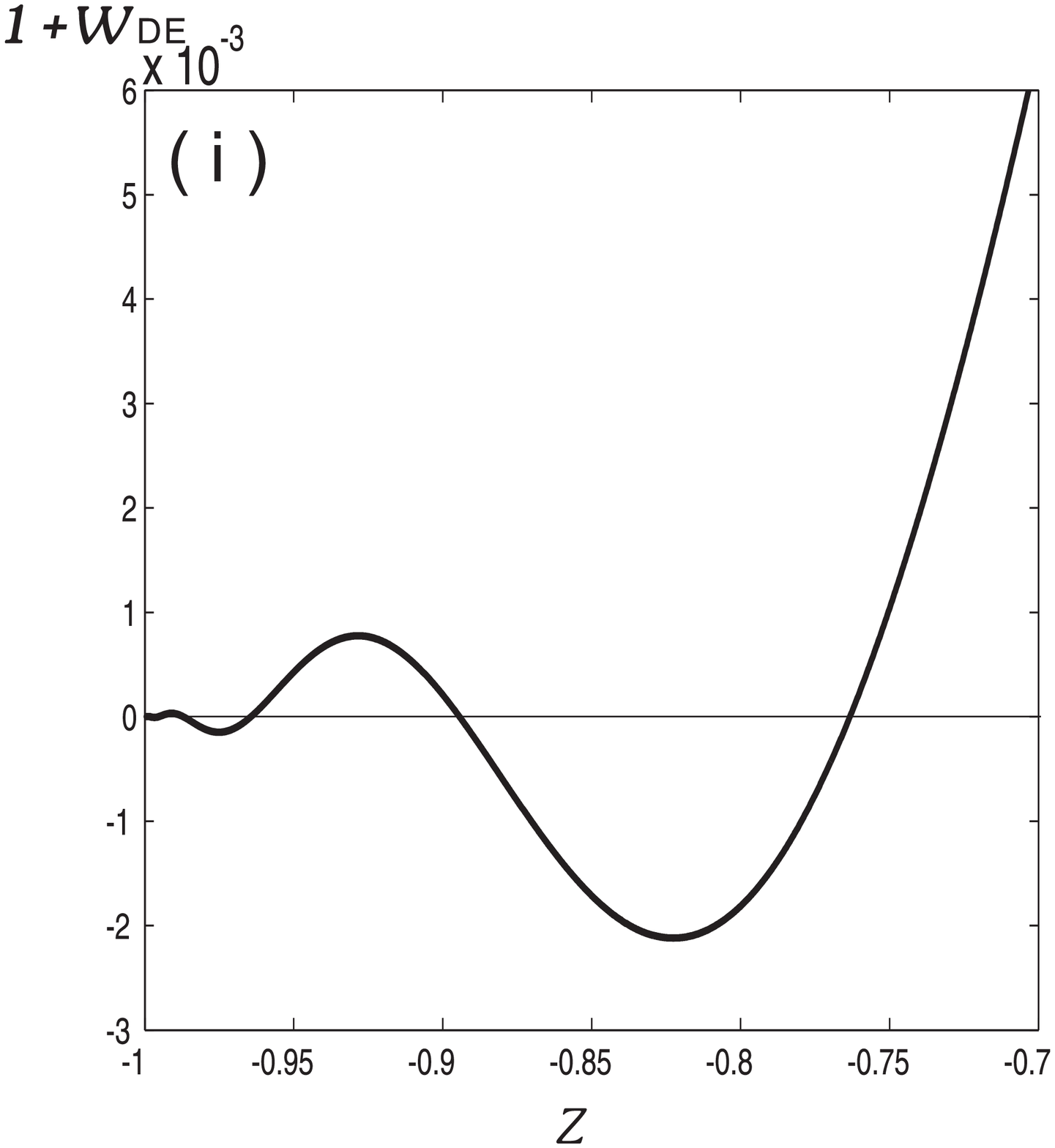}
                  }
\end{center}
\end{minipage}
&
\begin{minipage}{80mm}
\begin{center}
\unitlength=1mm
\resizebox{!}{6.5cm}{
   \includegraphics{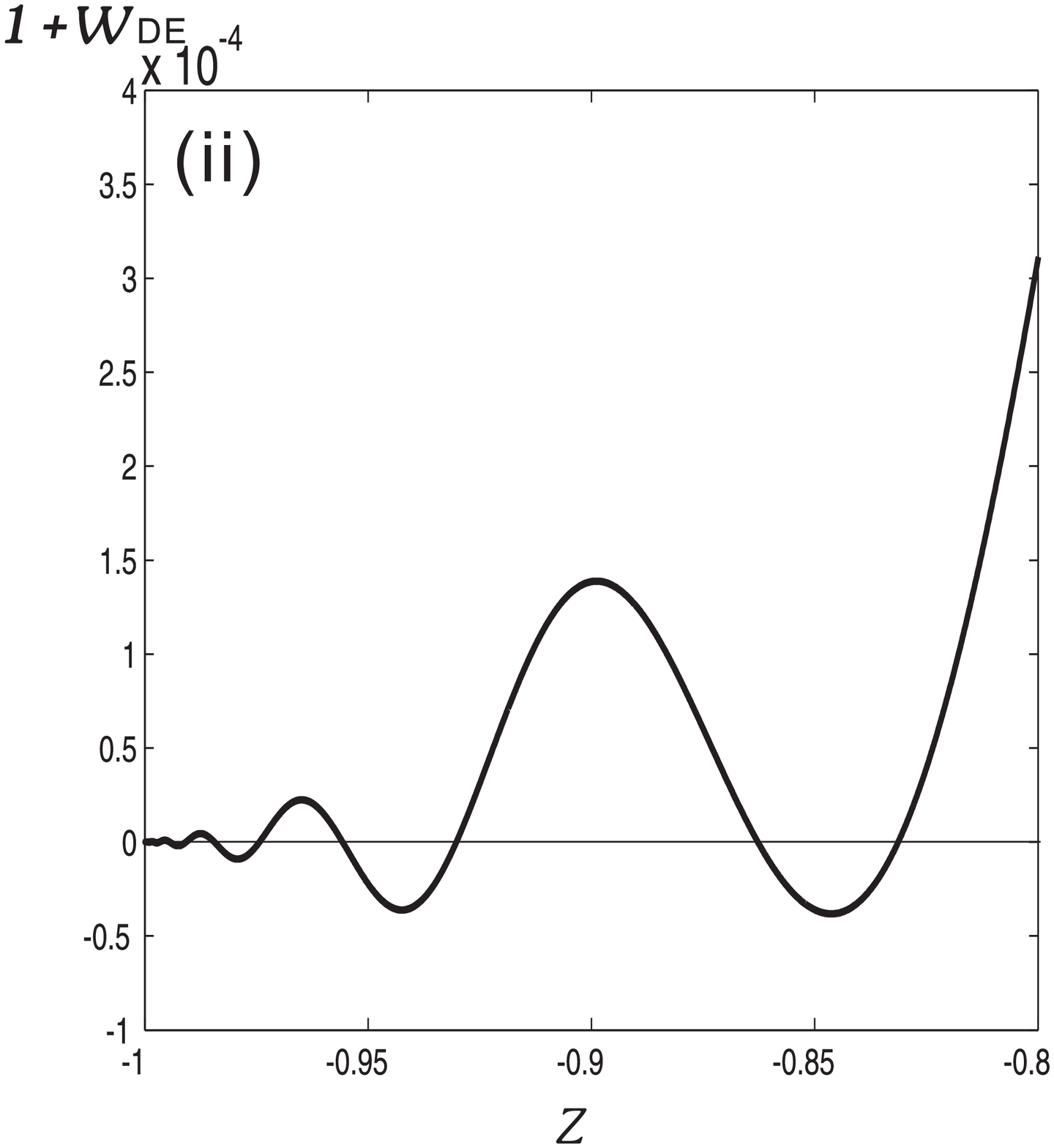}
                  }
\end{center}
\end{minipage}\\[5mm]
\begin{minipage}{80mm}
\begin{center}
\unitlength=1mm
\resizebox{!}{6.5cm}{
   \includegraphics{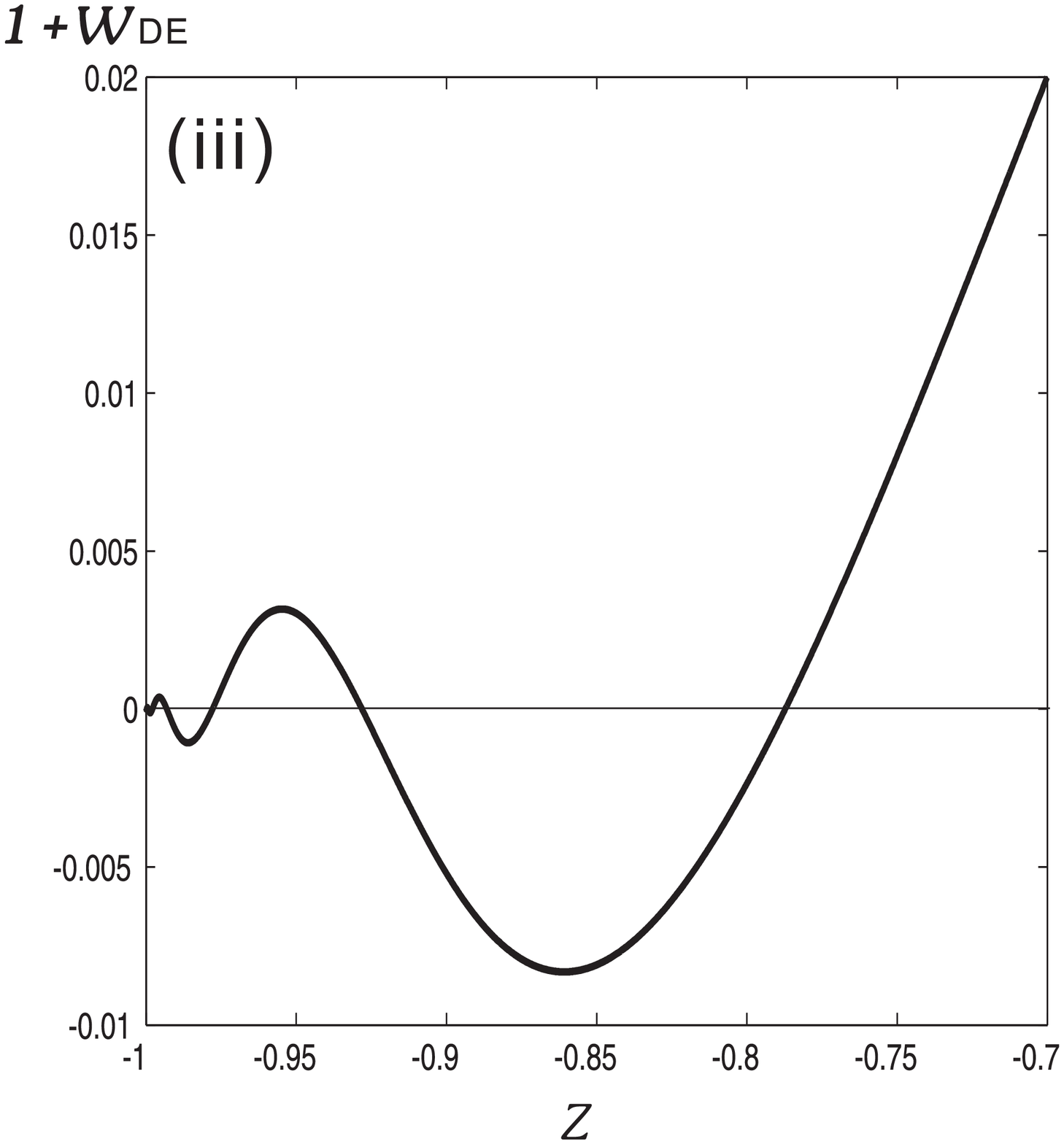}
                  }
\end{center}
\end{minipage}
&
\begin{minipage}{80mm}
\begin{center}
\unitlength=1mm
\resizebox{!}{6.5cm}{
   \includegraphics{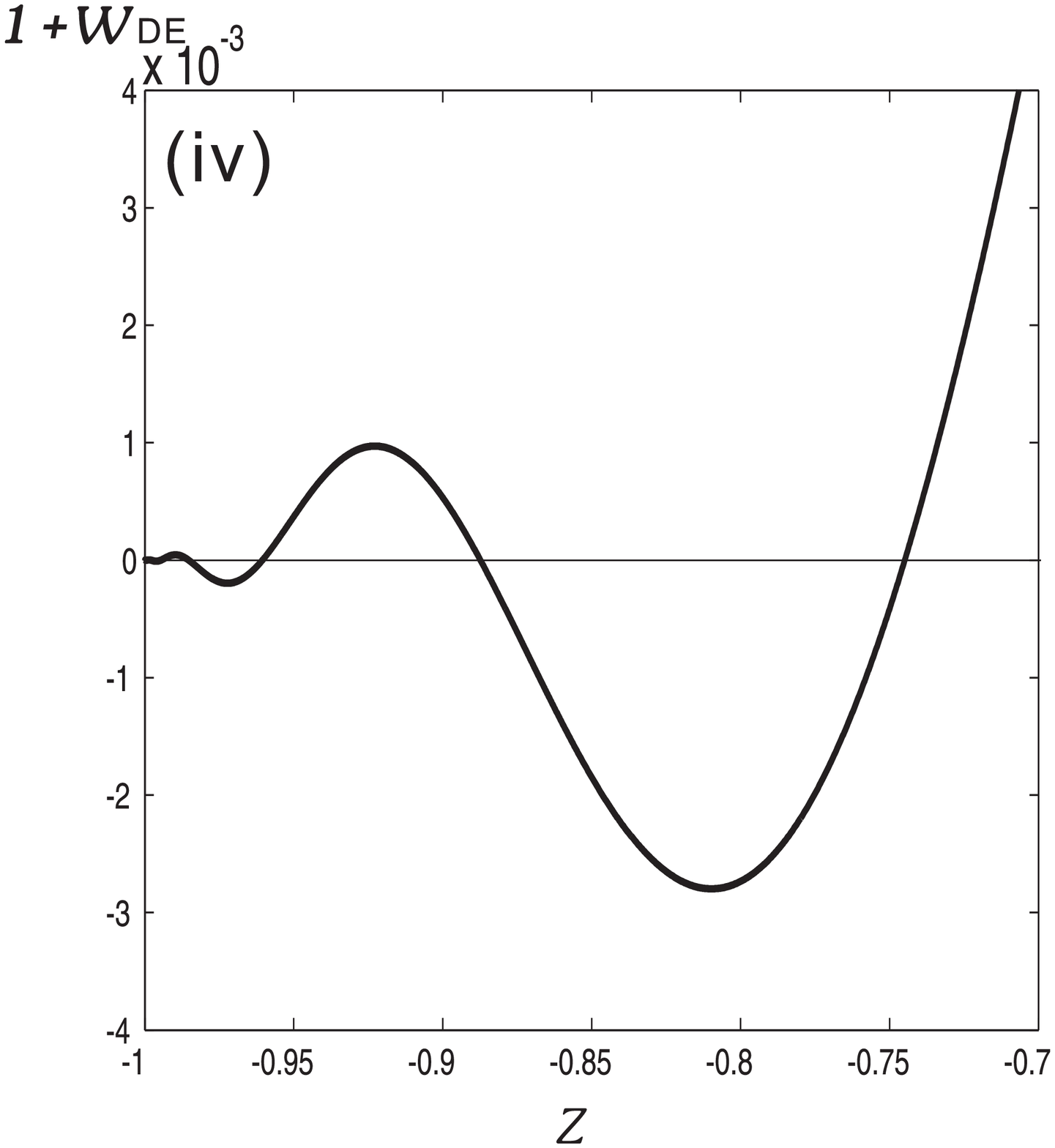}
                  }
\end{center}
\end{minipage}
%\\[5mm]
%\begin{minipage}{80mm}
%\begin{center}
%\unitlength=1mm
%\resizebox{!}{6.5cm}{
%   \includegraphics{Exponential-w.eps}
%                  }
%\end{center}
%\end{minipage}
%&
%\begin{minipage}{80mm}
%\begin{center}
%\unitlength=1mm
%\end{center}
%\end{minipage}
\end{tabular}
\caption{Future evolutions of 
$1+w_{\mathrm{DE}}$ 
as functions of the redshift $z$ 
in 
(i) Hu-Sawicki model for $p=1$, $c_1=2$ and $c_2=1$, 
(ii) Starobinsky model for $n=2$ and $\lambda=1.5$, 
%(iii) Appleby-Battye model for $b=2$, 
(iii) Tsujikawa model for $\mu=1$
and 
(iv) the exponential gravity model for $\beta=1.8$, 
respectively. 
The thin solid lines show $1+w_{\mathrm{DE}}=0$ (cosmological constant). 
}
\label{fg:1}
\end{figure}
\end{center}
%%%%%%%%%%%%%%%%%%%%%%

%%%%%% Fig. 2 %%%%%%%%%
\begin{center}
\begin{figure}[tbp]
\begin{tabular}{ll}
\begin{minipage}{80mm}
\begin{center}
\unitlength=1mm
\resizebox{!}{6.5cm}{
   \includegraphics{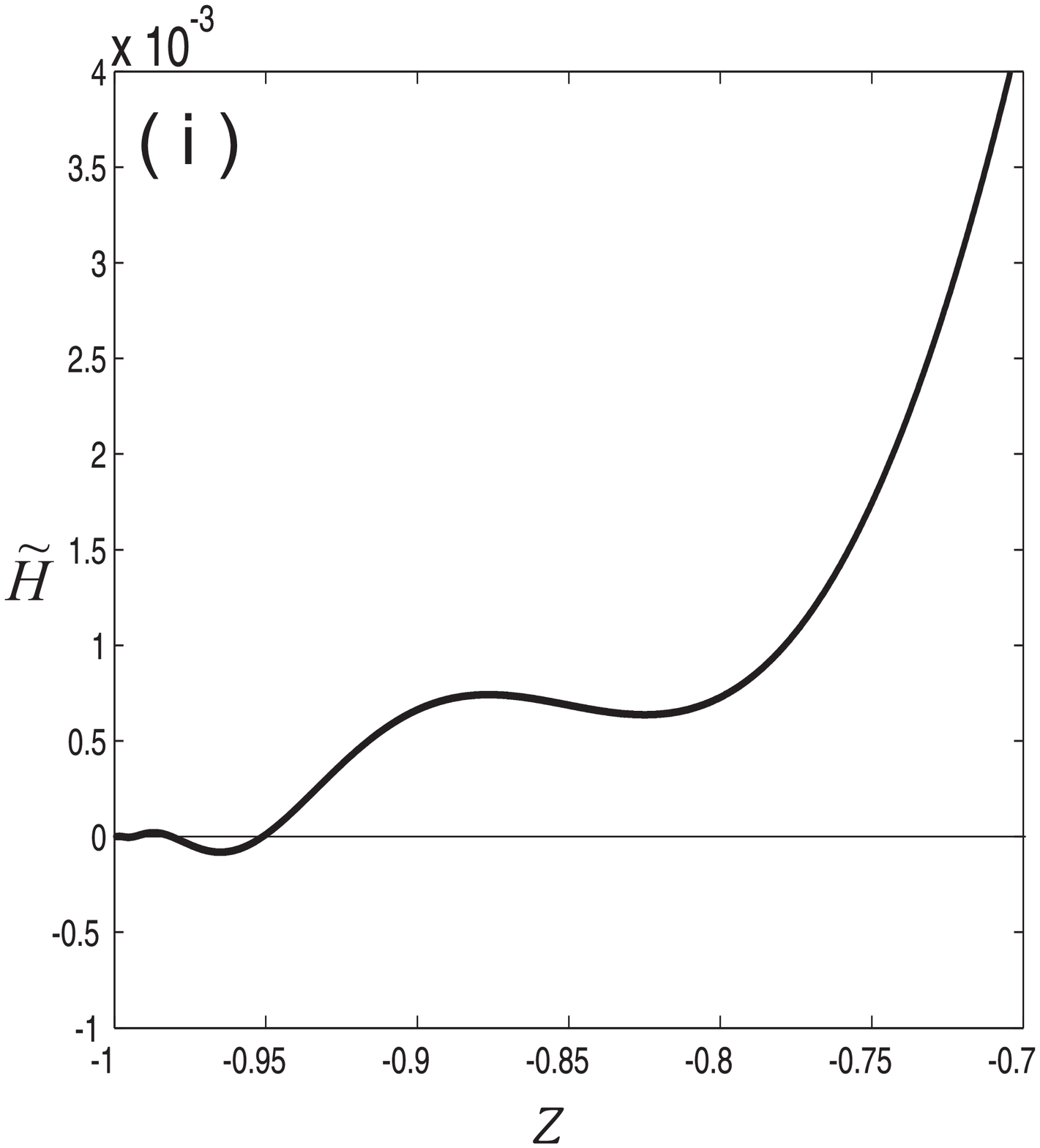}
                  }
\end{center}
\end{minipage}
&
\begin{minipage}{80mm}
\begin{center}
\unitlength=1mm
\resizebox{!}{6.5cm}{
   \includegraphics{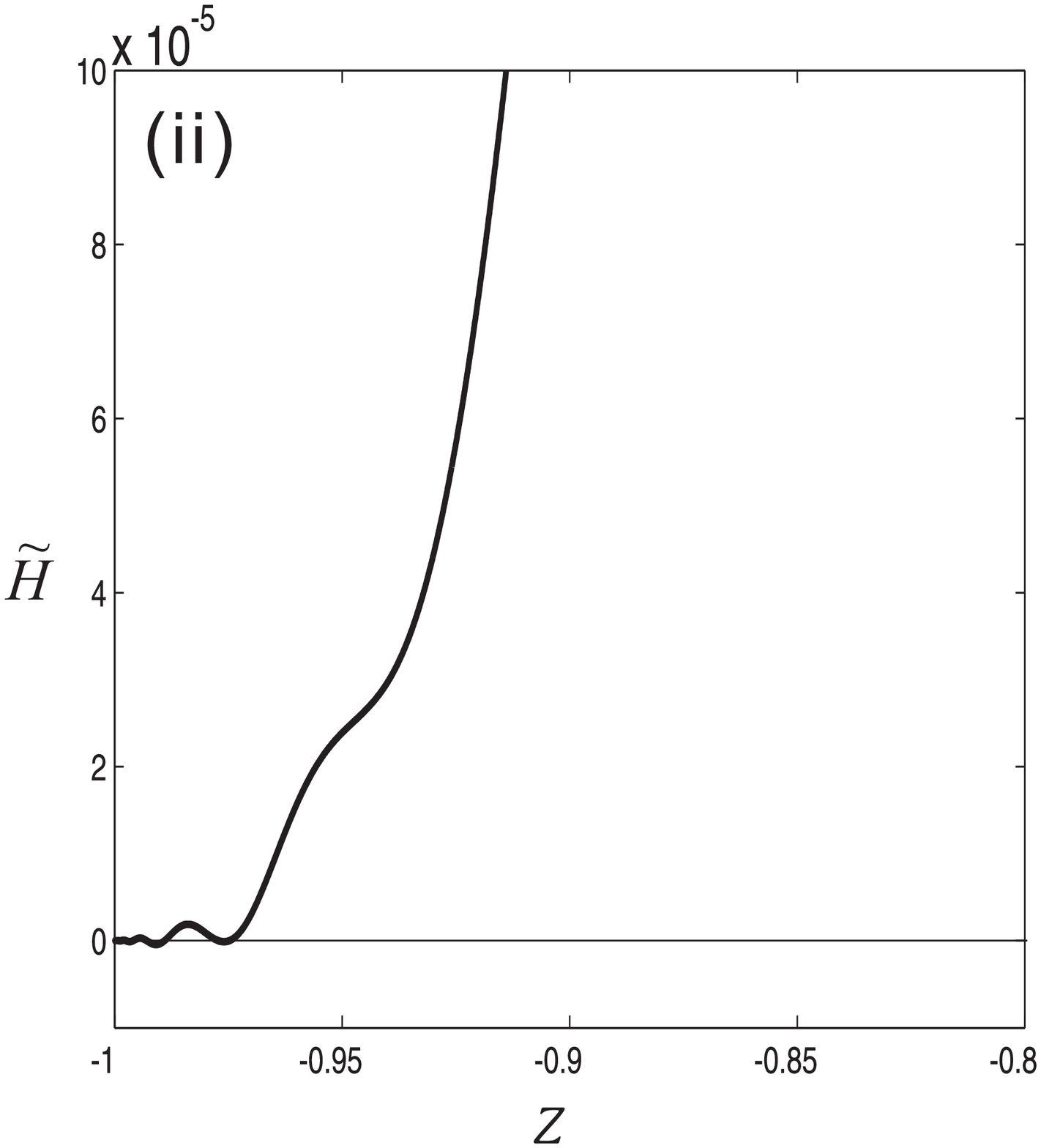}
                  }
\end{center}
\end{minipage}\\[5mm]
\begin{minipage}{80mm}
\begin{center}
\unitlength=1mm
\resizebox{!}{6.5cm}{
   \includegraphics{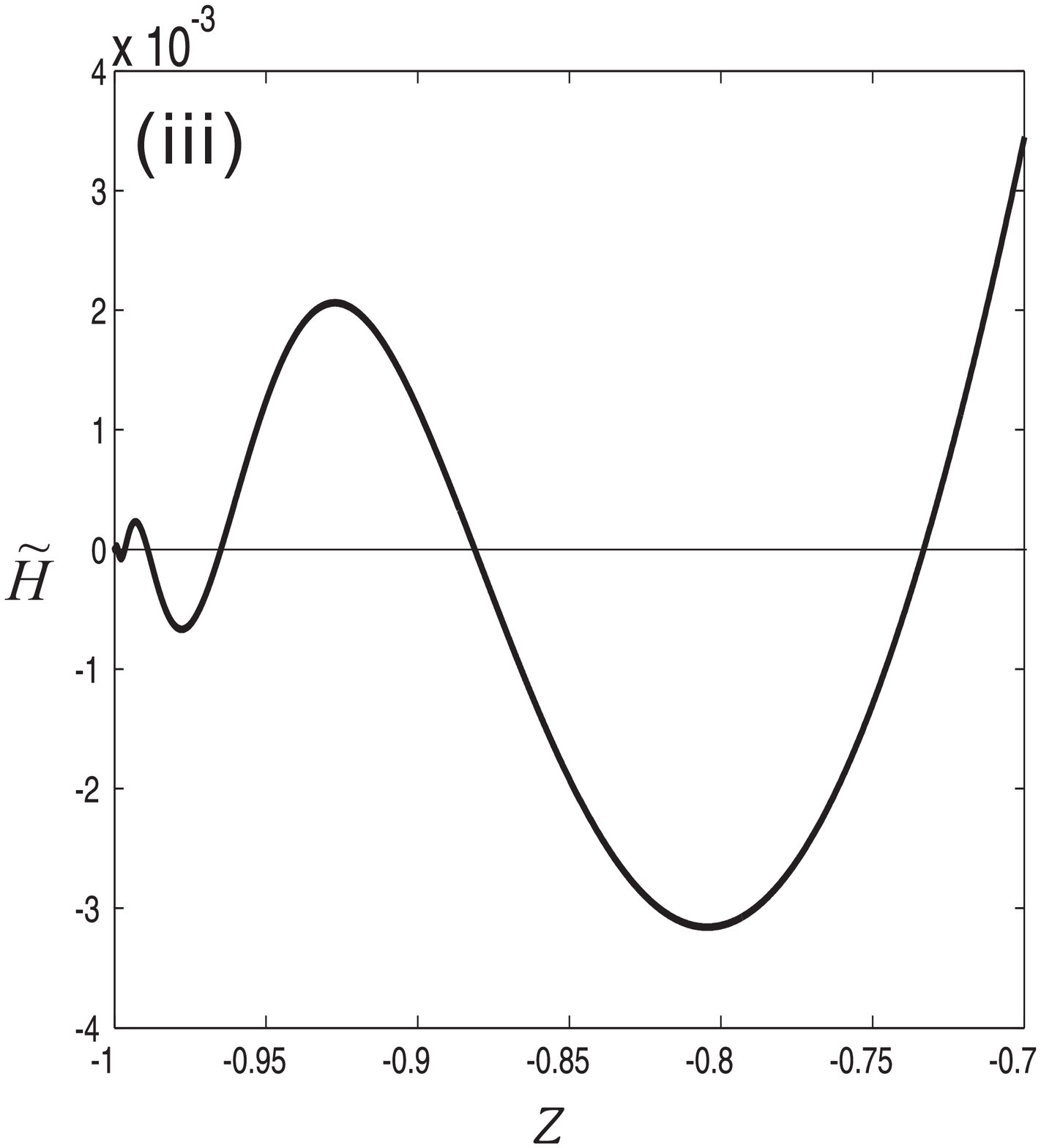}
                  }
\end{center}
\end{minipage}
&
\begin{minipage}{80mm}
\begin{center}
\unitlength=1mm
\resizebox{!}{6.5cm}{
   \includegraphics{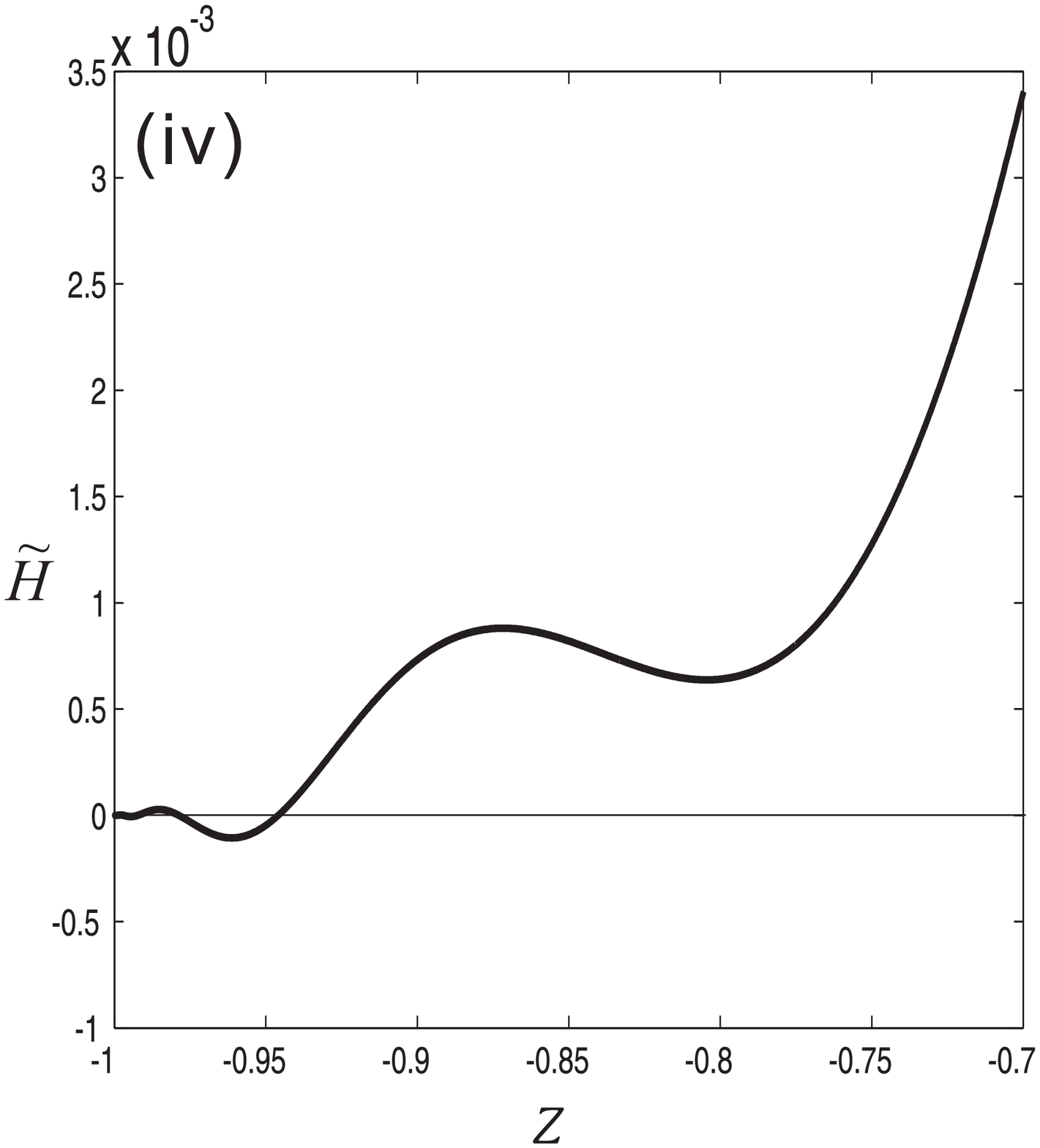}
                  }
\end{center}
\end{minipage}
%\\[5mm]
%\begin{minipage}{80mm}
%\begin{center}
%\unitlength=1mm
%\resizebox{!}{6.5cm}{
%   \includegraphics{Exponential-H.eps}
%                  }
%\end{center}
%\end{minipage}
%&
%\begin{minipage}{80mm}
%\begin{center}
%\unitlength=1mm
%\end{center}
%\end{minipage}
\end{tabular}
\caption{Future evolutions of 
$\tilde{H} \equiv \bar{H} - \bar{H}_{\mathrm{f}}$ 
with $\bar{H} \equiv H/H_0$ and $\bar{H}_{\mathrm{f}} \equiv H(z=-1)/H_0$ 
as functions of the redshift $z$. 
%The thin solid lines show $\tilde{H}=0$. 
Legend is the same as Fig.~1. 
}
\label{fg:2}
\end{figure}
\end{center}
%%%%%%%%%%%%%%%%%%%%%%

%%%%%% Fig. 3 %%%%%%%%%
\begin{center}
\begin{figure}[tbp]
\begin{tabular}{ll}
\begin{minipage}{80mm}
\begin{center}
\unitlength=1mm
\resizebox{!}{6.5cm}{
   \includegraphics{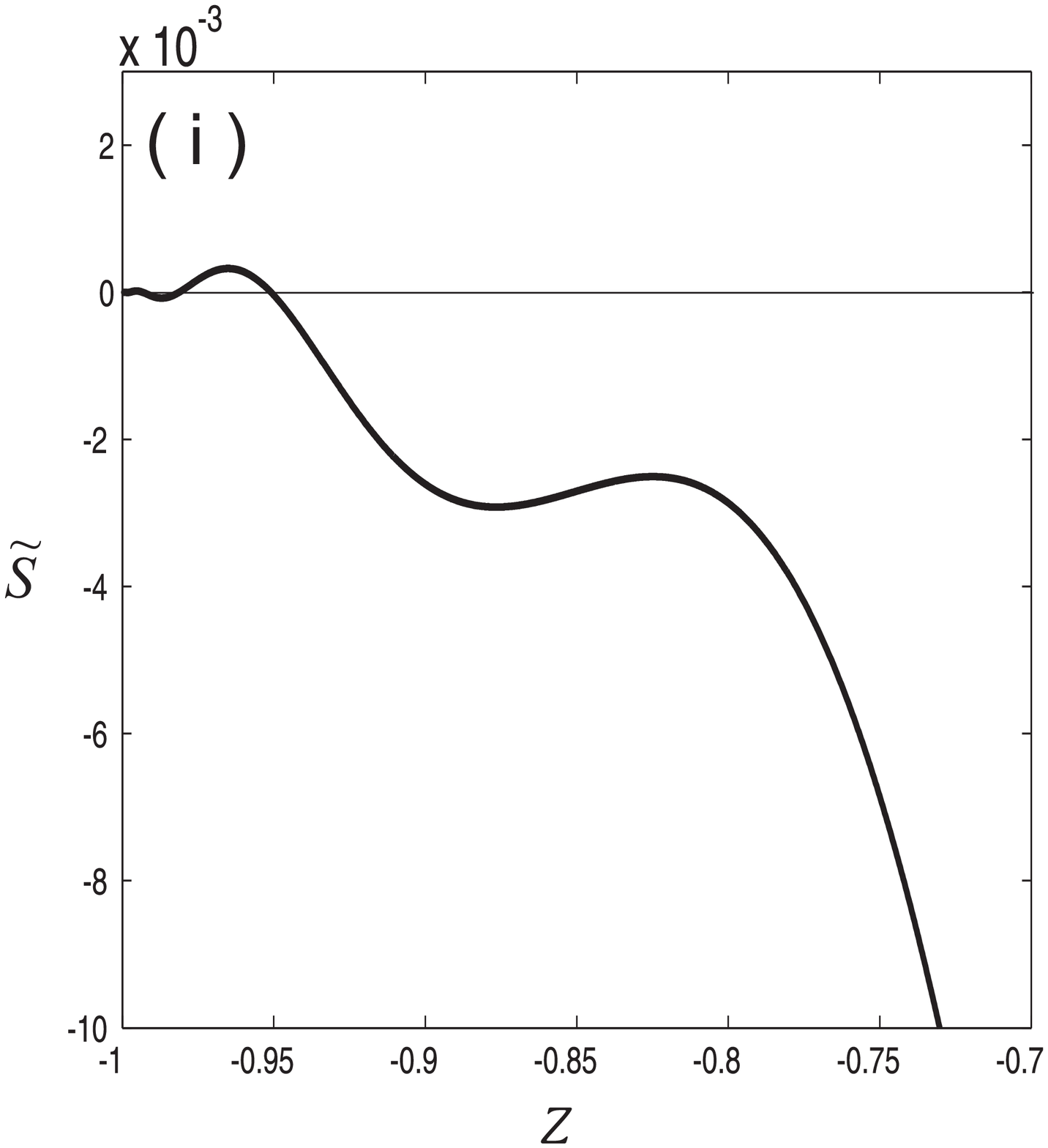}
                  }
\end{center}
\end{minipage}
&
\begin{minipage}{80mm}
\begin{center}
\unitlength=1mm
\resizebox{!}{6.5cm}{
   \includegraphics{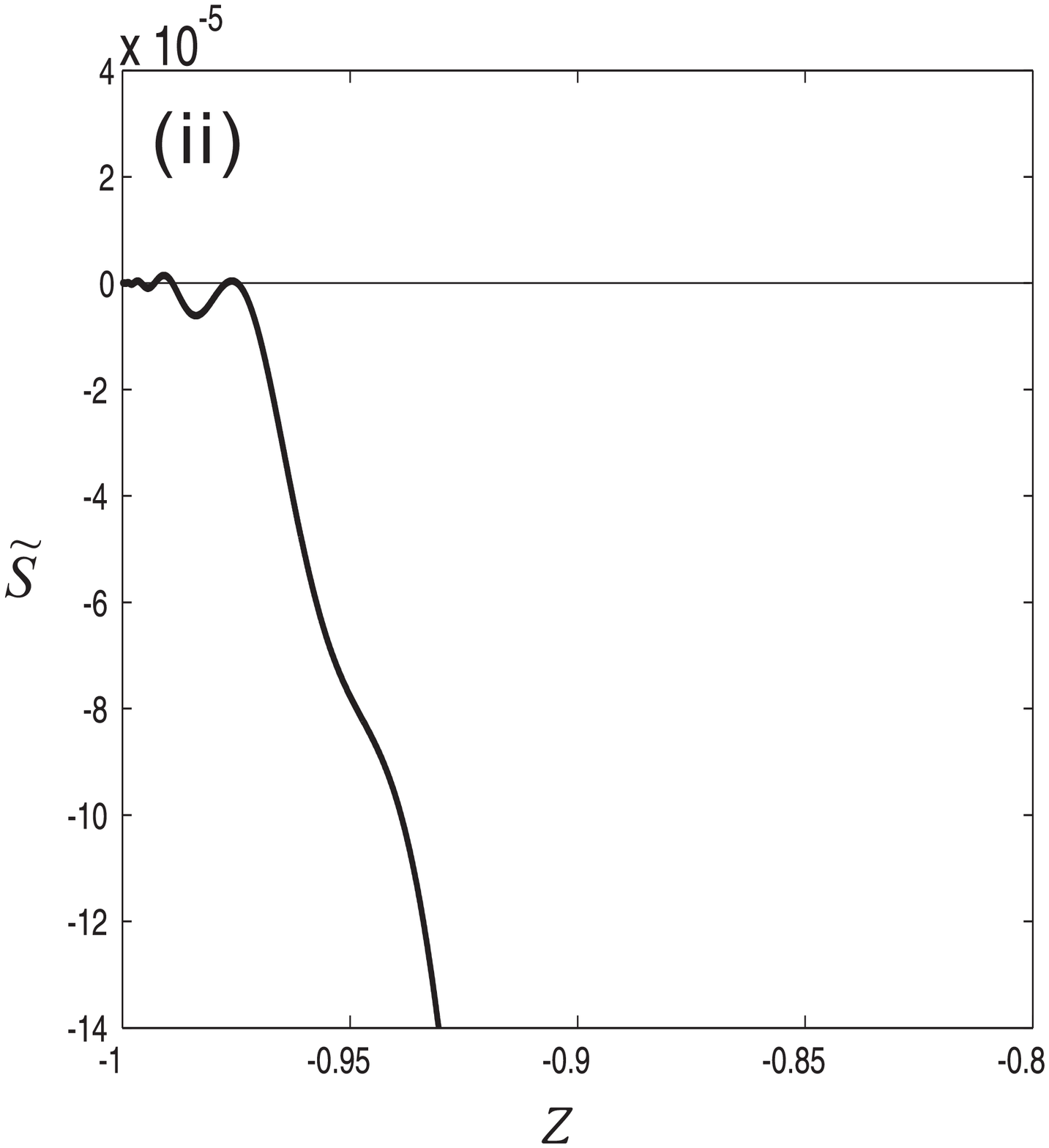}
                  }
\end{center}
\end{minipage}\\[5mm]
\begin{minipage}{80mm}
\begin{center}
\unitlength=1mm
\resizebox{!}{6.5cm}{
   \includegraphics{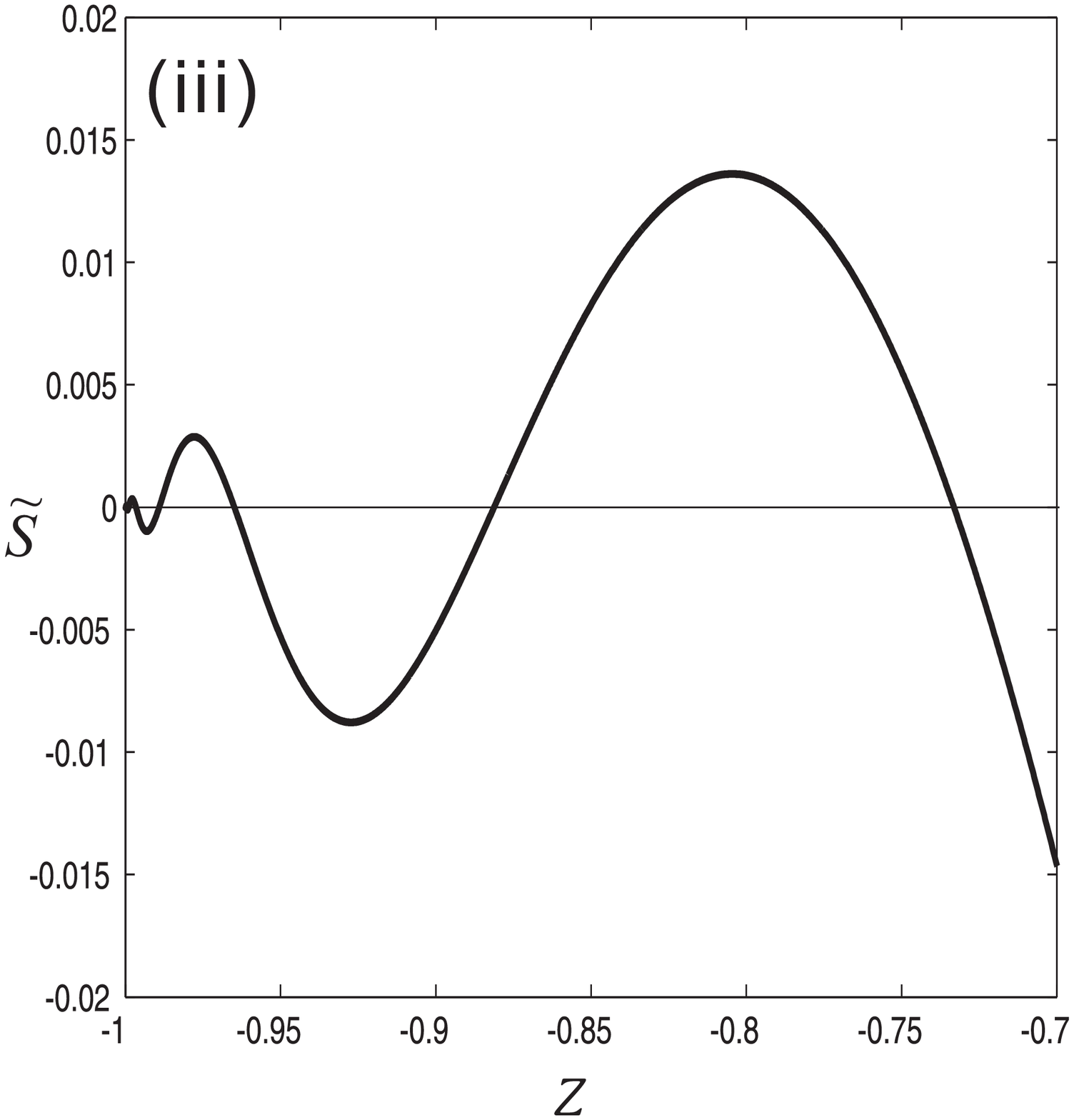}
                  }
\end{center}
\end{minipage}
&
\begin{minipage}{80mm}
\begin{center}
\unitlength=1mm
\resizebox{!}{6.5cm}{
   \includegraphics{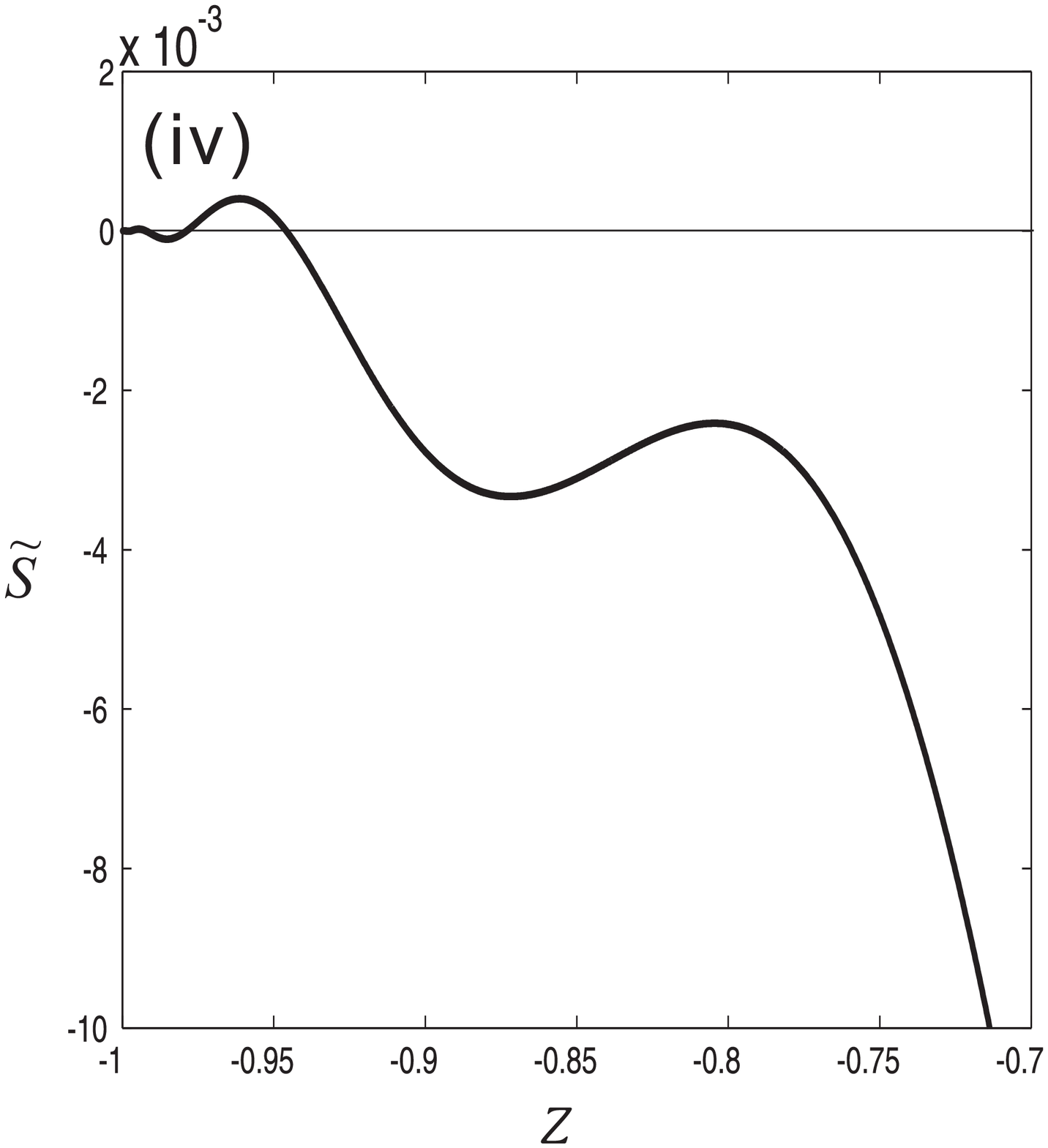}
                  }
\end{center}
\end{minipage}
%\\[5mm]
%\begin{minipage}{80mm}
%\begin{center}
%\unitlength=1mm
%\resizebox{!}{6.5cm}{
%   \includegraphics{Exponential-S.eps}
%                  }
%\end{center}
%\end{minipage}
%&
%\begin{minipage}{80mm}
%\begin{center}
%\unitlength=1mm
%\end{center}
%\end{minipage}
\end{tabular}
\caption{Future evolutions of 
$\tilde{S} \equiv \bar{S} - \bar{S}_{\mathrm{f}}$ 
with $\bar{S} \equiv S/S_0$ and $\bar{S}_{\mathrm{f}} \equiv S(z=-1)/S_0$ 
as functions of the redshift $z$. 
%The thin solid lines show $\tilde{S}=0$. 
Legend is the same as Fig.~1. 
}
\label{fg:3}
\end{figure}
\end{center}
%%%%%%%%%%%%%%%%%%%%%%

%%%%%%%%%%%%%%
\vspace{-18mm}
%%%%%%%%%%%%%%

In Figs.~1, 2 and 3, 
we depict the future evolutions of $1+w_{\mathrm{DE}}$, 
$\tilde{H} \equiv \bar{H} - \bar{H}_{\mathrm{f}}$ 
with $\bar{H} \equiv H/H_0$ and $\bar{H}_{\mathrm{f}} \equiv H(z=-1)/H_0$, 
and 
$\tilde{S} \equiv \bar{S} - \bar{S}_{\mathrm{f}}$ 
with $\bar{S} \equiv S/S_0$ and $\bar{S}_{\mathrm{f}} \equiv S(z=-1)/S_0$ 
as functions of the redshift $z \equiv 1/a -1$ 
in 
(i) Hu-Sawicki model for $p=1$, $c_1=2$ and $c_2=1$, 
(ii) Starobinsky model for $n=2$ and $\lambda=1.5$, 
(iii) Tsujikawa model for $\mu=1$
and 
(iv) the exponential gravity model for $\beta=1.8$, 
respectively. 
Here, $S_0 = \pi/\left(GH_0^2\right)$ is the present value of the horizon 
entropy $S$, $H_0$ is the current Hubble parameter, 
and the subscript `f' denotes the value at the final stage $z=-1$. 
Note that the present time is $z=0$ and the future is $-1 \leq z<0$. 
The parameters used for each model 
in Figs.~1--3 are the viable 
ones~\cite{Capozziello:2007eu, Tsujikawa:2009ku}. 
Several remarks are as follows:  
(a) the qualitative results do not strongly depend on the values of 
the parameters in each model; (b)
the evolutions of the Wald entropy $\hat{S}$ are similar to 
$S$~\cite{Bamba:2010ws, Bamba:2009id} in the models of (i)--(iv);
and (c) we have studied the Appleby-Battye model~\cite{Appleby:2007vb}, which 
is also a viable $f(R)$ model, and we have found that the numerical results 
are similar to those in the Starobinsky model of (ii) as expected.

%%%%%
We note that the present values of  
$w_{\mathrm{DE}}(z=0)$ are  -0.92, -0.97, -0.92 and -0.93 
for the models of (i)--(iv), respectively. 
These values satisfy the present observational 
constraints~\cite{Komatsu:2010fb}. 
%%%%%%%%
%%%%%%%%
Moreover, 
a dimensionless 
quantity $H^2/\left(\kappa^2 \rho_{\mathrm{m}}^{(0)}/3\right)$ 
can be determined through the numerical calculations, 
where $\rho_{\mathrm{m}}^{(0)}$ is the energy density of 
non-relativistic matter at the present time. 
If we use the observational data on the current density parameter of 
non-relativistic matter 
$\Omega_{\mathrm{m}}^{(0)} \equiv 
\rho_{\mathrm{m}}^{(0)}/\rho_{\mathrm{crit}}^{(0)}
= 0.26$ with 
$\rho_{\mathrm{crit}}^{(0)} = 3H_0^2/\kappa^2$~\cite{Komatsu:2010fb}, 
we find that the present value of the Hubble parameter 
$H_0 = H(z=0)$ is $H_0 = 71 \mathrm{km/s/Mpc}$~\cite{Komatsu:2010fb} 
for all the models of (i)--(iv). 
%%%%%%%%
%%%%%%%%
Furthermore, 
$(\bar{H}_{\mathrm{f}}, \bar{S}_{\mathrm{f}})  = 
(0.80, 1.6)$, $(0.85, 1.4)$, $(0.78, 1.7)$ and $(0.81, 1.5)$, 
for the models of (i)--(iv), respectively,
where $\bar{H}_{\mathrm{f}} \equiv H(z=-1)/H_0$ 
and $\bar{S}_{\mathrm{f}} \equiv S(z=-1)/S_0$. 
%%%%%

It is clear from Figs.~1--3 that 
in the future ($-1 \leq z \lesssim -0.74$), 
the crossings of the phantom divide 
are the generic feature for all the existing viable $f(R)$ models. 
%%%%%
By writing the first future crossing of the phantom divide 
and 
the first sign change of $\dot{H}$ from 
negative to positive as
$z = z_{\mathrm{cross}}$ 
and $z = z_{\mathrm{p}}$, respectively, we find that
 $(z_{\mathrm{cross}}, z_{\mathrm{p}})_{\alpha} = (-0.76, -0.82)_i$, 
$(-0.83, -0.98)_{ii}$, 
$(-0.79, -0.80)_{iii}$ and $(-0.74, -0.80)_{iv}$,
where the subscript $\alpha$ represents the $\alpha$th viable model. 
The values of the ratio 
$\Xi \equiv \Omega_{\mathrm{m}}/\Omega_{\mathrm{DE}}$ at 
$z = z_{\mathrm{cross}}$ and $z = z_{\mathrm{p}}$ are 
$(\Xi(z = z_{\mathrm{cross}}), \Xi(z = z_{\mathrm{p}}))_{\alpha} = 
(5.2 \times 10^{-3}, 2.1 \times 10^{-3})_i$, 
$(1.7 \times 10^{-3}, 4.8 \times 10^{-6})_{ii}$, 
$(4.1 \times 10^{-3}, 3.1 \times 10^{-3})_{iii}$ and 
$(6.2 \times 10^{-3}, 2.8 \times 10^{-3})_{iv}$, 
where 
$\Omega_{\mathrm{DE}} \equiv \rho_{\mathrm{DE}}/\rho_{\mathrm{crit}}^{(0)}$ 
and 
$\Omega_{\mathrm{m}} \equiv \rho_{\mathrm{m}}/\rho_{\mathrm{crit}}^{(0)}$ 
are the density parameters of dark energy and non-relativistic matter 
(cold dark matter and baryon), 
respectively. 
%
%and $\rho_{\mathrm{crit}}^{(0)} = 3H_0^2/\kappa^2$ 
%is the critical density. 
%
%%%%%
%%%%%
As $z$ decreases ($-1 \leq z \lesssim -0.90$), 
dark energy becomes much more 
dominant over non-relativistic matter 
($\Xi = \Omega_{\mathrm{m}}/\Omega_{\mathrm{DE}} \lesssim 10^{-5}$). 
As a result, one has $w_{\mathrm{DE}} \approx 
w_{\mathrm{eff}} \equiv 
-1 -2\dot{H}/\left(3H^2\right) 
= P_{\mathrm{tot}}/\rho_{\mathrm{tot}} 
$, 
where $w_{\mathrm{eff}}$ is the effective equation of state for the universe, 
and 
$\rho_{\mathrm{tot}} \equiv \rho_{\mathrm{DE}} + \rho_{\mathrm{m}} + 
\rho_{\mathrm{r}}$ and 
$P_{\mathrm{tot}} \equiv P_{\mathrm{DE}}  + P_{\mathrm{r}}$ 
are the total energy density and pressure of the universe, 
respectively. Here, $\rho_{\mathrm{m(r)}}$ 
and 
$P_{\mathrm{r}}$ are 
the energy density of non-relativistic matter (radiation) and 
the pressure of 
radiation, respectively. 
The physical reason why the crossing of the phantom divide appears 
in the farther future ($-1 \leq z \lesssim -0.90$) is that 
the sign of $\dot{H}$ 
changes from negative to positive due to the dominance of dark energy over 
non-relativistic matter. 
%%%%%
As $w_{\mathrm{DE}} \approx w_{\mathrm{eff}}$ in the farther future, 
$w_{\mathrm{DE}}$ oscillates around the phantom divide line 
$w_{\mathrm{DE}}=-1$ because the sign of $\dot{H}$ changes 
and consequently multiple crossings 
can be realized. 
We remark that since $S \propto H^{-2}$, 
the oscillating behavior of $S$ comes from that of $H$. 
However, it should be emphasized  that although
$S$ decreases in some regions, 
the second law of thermodynamics in $f(R)$ gravity 
can be always satisfied because $S$ is the cosmological horizon entropy
and it is not the total entropy of the universe including the entropy of 
generic matter. 
It has been shown that the second law of thermodynamics can be verified in 
both phantom and non-phantom phases for the same temperature of 
the universe outside and inside the apparent horizon 
in Ref.~\cite{Bamba:2010kf}. 

Finally, we mention that in our numerical calculations, 
we have taken the initial conditions  of   $z_0=8.0$, $8.0$, $3.0$ 
and $3.5$ for the models of (i)--(iv)
at $z=z_0$, respectively, so that 
$RF^{\prime}(z=z_0) \sim 10^{-13}$ with $F^{\prime}=dF/dR$,
to ensure that they
 can be all close enough to the $\Lambda\mathrm{CDM}$ model
with $RF^{\prime} =0$. 
We note that in order to save the calculation time, 
the different values of $z_0$  mainly reflect
the forms of the models, i.e., 
the power-law types of (i) and (ii) and 
 the exponential ones of  (iii) and (iv). 
It is clear that the smaller $RF^{\prime}(z=z_0)$ is, the closer the model to
$\Lambda\mathrm{CDM}$. However, the cut off of $RF^{\prime}(z=z_0) \sim 10^{-13}$
is assumed to evade the divergence of the calculations in the computing program.
We have also checked 
the results under the initial conditions with 
$R$, $f$, $F$, and $F^{\prime}$ as the $\Lambda\mathrm{CDM}$ values 
at the same redshift up to $z_0 = 4.0$ in the models of (i)--(iv) and found 
that they are qualitatively similar. 
We remark that at $z=z_{0}$, 
$w_{\mathrm{DE}} = -1$. 
%%%%%
%%%%%

%%%
Since $R/R_{\mathrm{c}} \gg 1$ in the high $z$ regime 
($z \simeq z_{0}$), 
the value of the combination $\gamma R_{\mathrm{c}}$ is set as 
$\gamma R_{\mathrm{c}} \simeq 18H_0^2 \Omega_{\mathrm{m}}^{(0)}$, 
where 
$(\gamma, R_{\mathrm{c}}) $ corresponds to $(c_1, R_{\mathrm{HS}})$, 
$(\lambda, R_{\mathrm{S}})$, 
$(\mu, R_{\mathrm{T}})$ 
and 
$(\beta, R_{\mathrm{E}})$ 
for 
(i) Hu-Sawicki ($c_2=1$), 
(ii) Starobinsky, 
(iii) Tsujikawa 
and 
(iv) the exponential gravity models, 
respectively, 
and $\Omega_{\mathrm{m}}^{(0)} \equiv 
\rho_{\mathrm{m}}^{(0)}/\rho_{\mathrm{crit}}^{(0)}
= 0.26$~\cite{Komatsu:2010fb}. 
%
%is the current density parameter of non-relativistic matter
%with $\rho_{\mathrm{m}}^{(0)}$ being the energy density of non-relativistic 
%matter at the present time. 
%
%%%%%
The reason is as follows. 
In the high $z$ regime ($z \simeq z_{0}$), $R/R_{\mathrm{c}} \gg 1$, in which 
$f(R)$ gravity has to be very close to the $\Lambda\mathrm{CDM}$ model,
 $\gamma R_{\mathrm{c}} \simeq 2 \Lambda = 
2\kappa^2 \left(\rho_{\mathrm{DE}}/\rho_{\mathrm{m}}^{(0)}\right) 
\left(\rho_{\mathrm{m}}^{(0)}/\rho_{\mathrm{crit}}^{(0)}\right) 
\rho_{\mathrm{crit}}^{(0)} 
=6\left(\rho_{\mathrm{DE}}/\rho_{\mathrm{m}}^{(0)}\right) H_0^2 
\Omega_{\mathrm{m}}^{(0)}$,
where $\Lambda$ is the effective cosmological constant in the limit of
$R/R_{\mathrm{c}} \gg 1$. 
As an initial condition, we take 
$\left(\rho_{\mathrm{DE}}/\rho_{\mathrm{m}}^{(0)}\right) 
= 3.0$~\cite{Bamba:2010ws}. 
Thus, we obtain 
$\gamma R_{\mathrm{c}} \simeq 18H_0^2 \Omega_{\mathrm{m}}^{(0)}$. 
%%%%%

%%%%%
Since the initial value at $z=z_0$ of the variable 
$y_H \equiv \rho_{\mathrm{DE}}/\rho_{\mathrm{m}}^{(0)}$ 
introduced in Ref.~\cite{Bamba:2010ws} is an arbitrary one, we have chosen it 
by using the observational data~\cite{Komatsu:2010fb} at the present time as 
$y_H \, (z=z_0) \simeq 
\Omega_{\mathrm{DE}}^{(0)}/\Omega_{\mathrm{m}}^{(0)}
\simeq 3.0$. 
The physical reason is as follows. 
By examining the cosmological evolutions of $y_H$ and $w_{\mathrm{DE}}$ 
as functions of the redshift $z$ for the models, 
we have found that 
$y_H \, (z=0)$ is close to its initial value of $y_H \, (z=z_0)$. 
This is because in the higher $z$ regime, 
the universe is in the phantom phase ($w_{\mathrm{DE}} < -1$) 
and therefore, $\rho_{\mathrm{DE}}$ and $y_H$ increase 
(since $y_H \propto \rho_{\mathrm{DE}}$), 
whereas in the lower $z$ regime, 
the universe is in the non-phantom (quintessence) phase 
($w_{\mathrm{DE}} > -1$) and hence they decrease. 
Consequently, the above two effects cancel out. 
%%%%%
%%%%%
%%%%%
To check our numerical results, 
by using the way in, e.g., Ref.~\cite{Tsujikawa:2009ku}, 
we have made another numerical calculation for the model of (iv) 
as follows. 
First, 
by taking a testing value of $\gamma R_{\mathrm{c}}$ at an initial value of 
$z_{0}$, 
we analyze the cosmological evolution numerically. 
Next, 
we find the adequate value of $z_{0}$ so that 
the current value of $y_H \, (z=0)$ can be equal to the present 
observational value of $\rho_{\mathrm{DE}}^{(0)}/\rho_{\mathrm{m}}^{(0)} 
= \left(1-\Omega_{\mathrm{m}}^{(0)}\right)/\Omega_{\mathrm{m}}^{(0)} 
= 2.85$, 
where $\rho_{\mathrm{DE}}^{(0)}$ is the present energy density of 
dark energy. 
We also obtain the initial value of $y_H \, (z=z_0)$ 
and the value of $R_{\mathrm{c}} 
\simeq 6 \gamma^{-1} y_H \, (z=z_0) \bar{m}^2$ 
with $\bar{m}^2 = H_0^2 \Omega_{\mathrm{m}}^{(0)}$. 
As a result, 
we have found that 
$y_H \, (z=z_0) = 2.72$  with $z_{0} = 3.0$ for $y_H \, (z=0) = 2.85$. 
Furthermore, 
we have confirmed that 
the obtained results with the above method are qualitatively similar to 
the ones shown in Figs. 1--3. 
Clearly, our results 
are not sensitive to the initial values of 
$z_0$ and $y_H \, (z=z_0)$. 
%%%%%
%%%%%
%%%%%
We note that the cosmological evolution of $w_{\mathrm{DE}}$ 
as a function of the redshift $z$ for the model of (iv) 
is given in Fig.~5 of Ref.~\cite{Bamba:2010ws}. 
On the other hand, 
the initial condition of 
$dy_H/d \ln a \, (z=z_0) = 0$ is due to 
that the $f(R)$ gravity models at $z=z_0$ should be very close to 
the $\Lambda\mathrm{CDM}$ model, in which $dy_H/d \ln a = 0$. 
%%%%%

%%%%%%%%%%%%%%%%%%%
%%%  Sec. IV
%%%%%%%%%%%%%%%%%%%
\section{Conclusions}

In the present paper, 
we have explored the future evolution of $w_{\mathrm{DE}}$ 
in the viable $f(R)$ gravity models and explicitly shown that in the future 
the crossings of the phantom divide 
are the generic feature in these models. 
We have also investigated the future evolution of the cosmological 
horizon entropy and demonstrated that the cosmological horizon entropy 
oscillates with time because the Hubble parameter also does. 
The new  cosmological ingredient obtained in this study is that in the future 
the sign of $\dot{H}$ changes from negative to positive due to the dominance 
of dark energy over non-relativistic matter. 
This is a common physical phenomena to the existing viable $f(R)$ 
models and thus it is one of the peculiar properties  of  $f(R)$ gravity 
models characterizing the deviation 
from the $\Lambda\mathrm{CDM}$ model.

%%%%%%%%%%%%%%%%%%%%%%%%
%%%  Acknowledgments
%%%%%%%%%%%%%%%%%%%%%%%%
\section*{Acknowledgments}

The work is supported in part by 
the National Science Council of R.O.C. under
Grant \#s: NSC-95-2112-M-007-059-MY3 and
NSC-98-2112-M-007-008-MY3
and 
National Tsing Hua University under the Boost Program and Grant \#: 
99N2539E1.

%%%%%%%%%%%%%%%%%%%%%%%%%%%%%%%%%
%% thebibliography environment
%%%%%%%%%%%%%%%%%%%%%%%%%%%%%%%%%

\end{document}